\documentclass[a5paper,twocolumn]{article}
\usepackage{graphicx}
\usepackage{amsmath,amssymb}
\makeatletter
\def\@maketitle{%
  \newpage
  \null
  \begin{center}%
  \let \footnote \thanks
    {\large\textbf \@title \par}%
    \vskip 1.5em%
    \lineskip .5em%
    \begin{tabular}[t]{c}%
      \@author
    \end{tabular}\par
    \vskip 1em%
    \@date
  \end{center}%
  \par
  \vskip 1.5em}
\renewcommand\section{%
  \@startsection {section}{1}{\z@}%
  {1\baselineskip \@plus 1ex \@minus .2ex}%
  {0.1ex \@plus.2ex}%
  {\normalfont\bfseries}}
\makeatother

\newcommand{\bnabla}{\boldsymbol{\nabla}}

\newcommand{\boldE}{\boldsymbol{E}}
\newcommand{\boldJ}{\boldsymbol{J}}
\newcommand{\boldB}{\boldsymbol{B}}
\newcommand{\boldq}{\boldsymbol{q}}

\begin{document}

\title{Numerical Computation of Thermoelectric and Thermomagnetic Effects}
\author{%
  H.~Okumura$^1$,
  S.~Yamaguchi$^2$,
  H.~Nakamura$^3$,
  K. Ikeda$^4$, and
  K. Sawada$^5$}

\date{%
  $^1$Matsusaka University, Matsusaka, 515-8511 Japan
  $\langle$\texttt{okumura@matsusaka-u.ac.jp}$\rangle$, \\
  $^2$National Institute for Fusion Science,
  $^3$Venture Business Laboratory, Nagoya University, \\
  $^4$Department of Fusion Science, Graduate University for Advanced Studies,
  $^5$Shinshu University}

\maketitle
\thispagestyle{empty}

\begin{abstract}
  Phenomenological equations describing the Seebeck, Hall, Nernst,
  Peltier, Ettingshausen, and Righi-Leduc effects are numerically
  solved for the temperature, electric current, and electrochemical
  potential distributions of semiconductors under magnetic field.  The
  results are compared to experiments.
\end{abstract}

\section{Introduction}

It is well known that geometry of samples (e.g., ratio of length to
width for rectangular samples) affects magnetoresistance.  Short
samples become more electrically resistant under magnetic field than
long ones (see Fig.~\ref{fig:hall} below).

For the Seebeck and the Nernst effects, experimental evidences for
such geometric contribution are not so clear.  Ertl~\cite{Ertl}
measured Bi-Sb alloy samples of various lengths, and showed that
longer samples exhibit greater Seebeck coefficients.  Measurement of
Seebeck coefficient by Ikeda \textit{et al}~\cite{Ikeda1}, however, is
not easy to summarize, but their Nernst coefficient of wider (``fat
bridge'') sample (Fig.~\ref{fig:fat}) under 4 Tesla was about 12
percent smaller than that of narrower one (Fig.~\ref{fig:thin}).

Though analytic solutions exist for a limited class of the Hall
effect~\cite{Wick}, numerical computation is necessary to explain
these results in general.  We developed a two-dimensional numerical
simulation code based on phenomenological equations governing
thermoelectric and thermomagnetic effects.  Section~\ref{eq:equations}
summarizes the basic equations.  Sections
\ref{sec:method}--\ref{sec:potential} describe the numerical
algorithm.  Sections \ref{sec:results} and \ref{sec:discussion}
summarizes the results and discuss consequences.

\section{Phenomenological Equations}
\label{eq:equations}

The phenomemological equations governing thermoelectric and
thermomagnetic phenomena of solids are~\cite{LL,HH}
\begin{gather}
  -\bnabla \phi
  = \rho \boldJ + \alpha\bnabla T
  + R \boldB \times \boldJ + N \boldB \times \bnabla T
  \label{eq:first} \\
  \boldq = \phi \boldJ - \kappa \bnabla T
  + \alpha T \boldJ + N T \boldB \times \boldJ
  + \kappa M \boldB \times \bnabla T
  \label{eq:second}
\end{gather}
where $\phi$ is the electrochemical potential per unit charge,\footnote{%
  For electrons, $\phi = \phi_{\text{electrostatic}} + \mu_n / (-e)$,
  where $\mu_n$ is the chemical potential of electrons
  with charge $-e < 0$.
  Similarly, $\phi = \phi_{\text{electrostatic}} + \mu_p / (+e)$
  for holes with charge $+e > 0$.
  If the system is close to equilibrium,
  the $\phi$'s for each species of carriers differ very little,
  so we shall not distinguish the $\phi$'s and
  the $\boldJ$'s for different carriers.}
$\rho$ the (isothermal) electric resistivity,
$\boldJ$ the electric current density,
$\alpha$ the (isothermal) Seebeck coefficient,
$T$ the temperature,
$R$ the (isothermal) Hall coefficient,
$\boldB$ the magnetic flux density,
$N$ the (isothermal transverse) Nernst coefficient,
$\boldq$ the energy flux density,
$\kappa$ the (isothermal) thermal conductivity,
and $M$ the Righi-Leduc coefficient.\footnote{%
  Note that the ``Leduc-Righi coefficient'' $L$ of Landau, Lifshitz,
  and Pitaevski\u{\i}~\cite{LL} corresponds to our $\kappa M$.}
In Eq.~(\ref{eq:first}), 
the last three terms represents
the Seebeck, the Hall, and the Nernst effects, respectively.
In Eq.~(\ref{eq:second}), 
the last three terms are responsible for
the Peltier/\linebreak[1]Thomson,
the Ettingshausen, and the Righi-Leduc effects, respectively.

In what follows we assume that
(1) the system is in steady state,
so that $\bnabla \cdot \boldJ = \bnabla \cdot \boldq = 0$ holds;
(2) the external magnetic field $\boldB$
is independent of the position,\footnote{%
  Rigorously speaking,
  external field $\boldB$ gives rise to electric current $\boldJ$,
  which in turn modifies the field $\boldB$
  according to the Maxwell equation
  $\bnabla \times \boldB = \mu \boldJ
  + \epsilon \mu \partial \boldE / \partial t$.
  (In steady state $\partial \boldE / \partial t = 0$.)
  In practice, however,
  magnetic permeability
  $\mu \sim \mu_0 = 4 \pi \times 10^{-7}$
  is so small that $\bnabla \times \boldB$
  can be safely neglected.}
and is along the $z$-direction;
(3) the electric current $\boldJ$ has no $z$-component;
(4) the temperature $T$ is independent of the $z$-coordinate;
(5) the conductor is homogenious,
so that all of the transport coefficients
($\alpha$, $\rho$, $\kappa$, $R$, $N$, $M$)
are functions of temperature $T$ alone.

\section{Overview of the Method}
\label{sec:method}

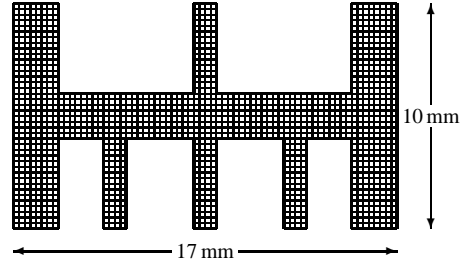
\begin{figure}
  \begin{center}
    \setlength{\unitlength}{0.75mm}
    \begin{picture}(80,40)(0,-4)
      \thicklines
      \put(0,0){\line(0,1){40}}
      \put(68,0){\line(0,1){40}}
      \put(0,0){\line(1,0){8}}
      \put(16,0){\line(1,0){4}}
      \put(32,0){\line(1,0){4}}
      \put(48,0){\line(1,0){4}}
      \put(60,0){\line(1,0){8}}
      \put(0,40){\line(1,0){8}}
      \put(32,40){\line(1,0){4}}
      \put(60,40){\line(1,0){8}}
      \put(8,0){\line(0,1){16}}
      \put(16,0){\line(0,1){16}}
      \put(20,0){\line(0,1){16}}
      \put(32,0){\line(0,1){16}}
      \put(36,0){\line(0,1){16}}
      \put(48,0){\line(0,1){16}}
      \put(52,0){\line(0,1){16}}
      \put(60,0){\line(0,1){16}}
      \put(8,24){\line(0,1){16}}
      \put(32,24){\line(0,1){16}}
      \put(36,24){\line(0,1){16}}
      \put(60,24){\line(0,1){16}}
      \put(8,16){\line(1,0){8}}
      \put(20,16){\line(1,0){12}}
      \put(36,16){\line(1,0){12}}
      \put(52,16){\line(1,0){8}}
      \put(8,24){\line(1,0){24}}
      \put(36,24){\line(1,0){24}}
      \thinlines
      \multiput(1,0)(1,0){7}{\line(0,1){40}}
      \multiput(8,16)(1,0){9}{\line(0,1){8}}
      \multiput(17,0)(1,0){3}{\line(0,1){24}}
      \multiput(20,16)(1,0){13}{\line(0,1){8}}
      \multiput(33,0)(1,0){3}{\line(0,1){40}}
      \multiput(36,16)(1,0){13}{\line(0,1){8}}
      \multiput(49,0)(1,0){3}{\line(0,1){24}}
      \multiput(52,16)(1,0){9}{\line(0,1){8}}
      \multiput(61,0)(1,0){7}{\line(0,1){40}}
      \multiput(0,17)(0,1){7}{\line(1,0){68}}
      \multiput(0,1)(0,1){16}{\line(1,0){8}}
      \multiput(0,24)(0,1){16}{\line(1,0){8}}
      \multiput(16,1)(0,1){16}{\line(1,0){4}}
      \multiput(32,1)(0,1){16}{\line(1,0){4}}
      \multiput(32,24)(0,1){16}{\line(1,0){4}}
      \multiput(48,1)(0,1){16}{\line(1,0){4}}
      \multiput(60,1)(0,1){16}{\line(1,0){8}}
      \multiput(60,24)(0,1){16}{\line(1,0){8}}
      \put(28,-4){\vector(-1,0){28}}
      \put(40,-4){\vector(1,0){28}}
      \put(34,-4){\makebox(0,0){\footnotesize 17\,mm}}
      \put(74,18){\vector(0,-1){18}}
      \put(74,22){\vector(0,1){18}}
      \put(74,20){\makebox(0,0){\footnotesize 10\,mm}}
    \end{picture}
  \end{center}
  \caption{Example of discretization.  This figure corresponds to
    the ``bridge'' shape of Ikeda \textit{et al}~\cite{Ikeda1}.}
  \label{fig:thinmesh}
\end{figure}

\begin{figure}
  \begin{center}
    \setlength{\unitlength}{1mm}
    \begin{picture}(80,50)
      \thicklines
      \put(20,25){\line(1,0){40}}
      \put(40,5){\line(0,1){40}}
      \thinlines
      \put(20,25){\circle*{1}}
      \put(40,5){\circle*{1}}
      \put(60,25){\circle*{1}}
      \put(40,45){\circle*{1}}
      \put(40,25){\circle*{1}}
      \put(41,24){\makebox(0,0)[tl]{$(i,j)$}}
      \put(20,24){\makebox(0,0)[t]{$(i-1,j)$}}
      \put(60,24){\makebox(0,0)[t]{$(i+1,j)$}}
      \put(41,5){\makebox(0,0)[l]{$(i,j-1)$}}
      \put(41,45){\makebox(0,0)[l]{$(i,j+1)$}}
      \put(48,26){\vector(1,0){4}}
      \put(50,27){\makebox(0,0)[b]{$J_{ij}^x$}}
      \put(28,26){\vector(1,0){4}}
      \put(30,27){\makebox(0,0)[b]{$J_{i-1,j}^x$}}
      \put(39,33){\vector(0,1){4}}
      \put(39,35){\makebox(0,0)[r]{$J_{ij}^y$}}
      \put(39,13){\vector(0,1){4}}
      \put(39,15){\makebox(0,0)[r]{$J_{i-1,j}^y$}}
    \end{picture}
  \end{center}
  \caption{Grid point $(i,j)$ and its four adjacent points.
    Temperature $T_{ij}$ and potential $\phi_{ij}$ are
    specified on point $(i,j)$,
    whereas electric current $J_{ij}^x$ is specified along the
    line segment connecting two points $(i,j)$ and $(i+1,j)$.}
  \label{fig:mesh}
\end{figure}
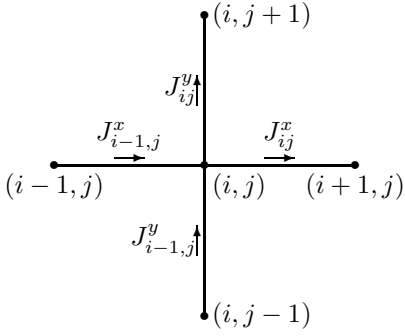

Our aim is to calculate $T$, $\phi$, and $\boldJ$ distributions for
rectangular and irregular-shaped semiconductor samples such as shown in
Fig.~\ref{fig:thinmesh}.
We discretize position by constructing a grid with square meshes of
size $h \times h$.  We consider $T$ and $\phi$ on each grid points
(corners of the meshes), and $\boldJ$ on each side of the meshes, as
shown in Fig.~\ref{fig:mesh}.

After setting up suitable initial values for $T$, $\phi$, and
$\boldJ$, we proceed as follows:
\begin{enumerate}
\item On each grid point $(i,j)$, update $T_{ij}$ from the discretized
  Poisson equation for temperature (see Section~\ref{sec:temp} below),
  assuming all the other quantities fixed.
\item On each line segment connecting adjacent grid points, update
  $\boldJ_{ij}$ (see Section~\ref{sec:current} below), assuming all
  the other quantities fixed.
\item On each grid point, update $\phi_{ij}$ so as to satisfy the
  continuity equation for $\boldJ_{ij}$ at the point
  (see Section~\ref{sec:potential} below).
\item Go to Step 1.
\end{enumerate}

\section{Temperature Updates}
\label{sec:temp}

The Poisson equation for temperature can be derived by taking the
divergence of Eq.~(\ref{eq:second}) and using Eq.~(\ref{eq:first}):
\begin{multline}
  \left( \kappa - \frac{N^2 T B^2}{\rho} \right) \nabla^2 T \\
  = - \rho J^2
  + \left( T \frac{d \alpha}{d T}
    + \frac{N T B^2}{\rho} \frac{d R}{d T}
  \right) (\bnabla T) \cdot \boldJ \\
  \shoveright{+ \left( 2N + T \frac{dN}{dT}
    - \frac{N T}{\rho} \frac{d \rho}{d T}
  \right) (\bnabla T) \cdot (\boldB \times \boldJ)} \\
  + \left(
    \frac{N T B^2}{\rho} \frac{d N}{d T} - \frac{d \kappa}{d T}
  \right) (\bnabla T)^2
  \label{eq:nabla2T}
\end{multline}

This second-order partial differential equation for $T$,
which we shall abbreviate as $\nabla^2 T(x, y) = F(x, y)$,
can be discretized as follows:
If the point $(i, j)$ is not on the boundary,
\begin{equation}
  T^\mathrm{new}_{ij} =
  \frac{T_{i-1,j} + T_{i+1,j} + T_{i,j-1} + T_{i,j+1}}{4}
  - \frac{h^2}{4} F
  \label{eq:updateT}
\end{equation}
(with suitable modification to accelerate convergence).
On the boundary, either $T_{ij}$ is given (Dirichlet conditions), or
derivatives of $T$ in the direction normal to the boundary, $\nabla_n
T(x,y)$, is given (Neumann conditions).  In the latter case, if the
grid point $(i,j)$ is on the boundary which is along the $x$-direction
such that point $(i-1,j)$ is outside of the sample, the normal
derivative $\nabla_y T = \partial T / \partial y$ should be given,
and the update formula is
\begin{equation}
  T^\mathrm{new}_{ij} =
  \frac{T_{i-1,j} + T_{i+1,j} + 2T_{i,j+1} - 2h \nabla_y T}{4}
  - \frac{h^2}{4} F
\end{equation}
Similarly, at a corner point such that
points $(i-1,j)$ and $(i,j-1)$ are outside of the boundary,
the update formula becomes
\begin{equation}
  T^\mathrm{new}_{ij} =
  \frac{2T_{i+1,j} + 2T_{i,j+1} - 2h \nabla_x T - 2h \nabla_y T}{4}
  - \frac{h^2}{4} F
\end{equation}

The normal derivative, say $\nabla_y T$, can be derived
from Eq.~(\ref{eq:second}) if there is no energy and current transfer
through the boundary ($q_y = J_y = 0$),
\begin{equation}
  \nabla_y T = N T B_z J_x / \kappa + M B_z \nabla_x T
  \label{eq:qy=jy=0}
\end{equation}

\section{Electric Current Calculation}
\label{sec:current}

Given $T$ and $\phi$, the electric current density $\boldJ$ can be
computed from Eq.~(\ref{eq:first}), which can be written as
\begin{equation}
  \rho \boldJ + R \boldB \times \boldJ
  = - \bnabla \phi - \alpha \bnabla T - N \boldB \times \bnabla T
\end{equation}
or, in coordinate components
\begin{equation}
  \begin{pmatrix}
    \rho & - R B_z \\
    R B_z & \rho
  \end{pmatrix}
  \begin{pmatrix}
    J_x \\ J_y 
  \end{pmatrix}
  =
  \begin{pmatrix}
    - \nabla_x \phi - \alpha \nabla_x T + N B_z \nabla_y T \\
    - \nabla_y \phi - \alpha \nabla_y T - N B_z \nabla_x T
  \end{pmatrix}
\end{equation}
Hence
\begin{multline}
  \begin{pmatrix}
    J_x \\ J_y 
  \end{pmatrix}
  =
  \frac{1}{\rho^2 + R^2 B_z^2}
  \begin{pmatrix}
    \rho & R B_z \\
    - R B_z & \rho
  \end{pmatrix} \\
  \times
  \begin{pmatrix}
    - \nabla_x \phi - \alpha \nabla_x T + N B_z \nabla_y T \\
    - \nabla_y \phi - \alpha \nabla_y T - N B_z \nabla_x T
  \end{pmatrix}
  \label{eq:JxJy}
\end{multline}

On the boundary, if $J_y = 0$ holds, then
\begin{equation}
  J_x = (- \nabla_x \phi - \alpha \nabla_x T + N B_z \nabla_y T) / \rho
  \label{eq:jy=0}
\end{equation}
Furthermore, if $q_y = 0$ also holds,
substitution of (\ref{eq:qy=jy=0}) into (\ref{eq:jy=0}) gives
\begin{equation}
  J_x = \frac{1}{\rho} \left(- \nabla_x \phi
    - ( \alpha - N M B_z^2 ) \nabla_x T
    + \frac{N^2 B_z^2 T J_x}{\kappa} \right)
\end{equation}
Solving for $J_x$, we have
\begin{equation}
  J_x =
  \left(
    - \nabla_x \phi
    - ( \alpha - N M B_z^2 ) \nabla_x T
  \right)
  \biggm/
  \left(
    \rho - \frac{N^2 B_z^2 T}{\kappa}
  \right)  \label{eq:jy=qy=0}
\end{equation}

\section{Potential Updates}
\label{sec:potential}

Instead of using a lengthy Poisson equation for the electrochemical
potential $\phi$ that can be derived from Eq.~(\ref{eq:first}), we
base our calculation of $\phi$ on the continuity equation of electric
current, $\bnabla \cdot \boldJ = 0$.

We start with Eq.~(\ref{eq:JxJy}) which has the form
\begin{multline}
  J_x = \frac{1}{\rho^2 + R^2 B^2}
  ( - \rho \nabla_x \phi - R B_z \nabla_y \phi ) \\
  + \text{terms independent of $\phi$}
\end{multline}
and similarly for $J_y$.
As was discussed earlier,
the discretized value $J^x_{ij}$ is taken
along the line segment connecting
two points with potentials $\phi_{ij}$ and $\phi_{i+1,j}$.
Along this line segment, we can approximate potential derivatives by
$\nabla_x \phi = (\phi_{i+1,j} - \phi_{ij}) / h$ and
$\nabla_y \phi = (\phi_{i,j+1} + \phi_{i+1,j+1}
- \phi_{i,j-1} - \phi_{i+1,j-1}) / (4h)$.
After these derivatives are substituted,
the equation for $J^x_{ij}$ above has the $\phi_{ij}$ dependence:
\begin{equation}
  J^x_{ij} = \frac{\rho \phi_{ij}}{h (\rho^2 + R^2 B^2)}
  + \text{terms independent of $\phi_{ij}$}
\end{equation}
Note that $J^x_{ij}$ is the current outgoing from the point $(i,j)$
in the positive $x$-direction.
As can be seen from Fig.~\ref{fig:mesh},
there are four such expressions outgoing from point $(i,j)$:
$J^x_{ij}$, $J^y_{ij}$, $-J^x_{i-1,j}$, $-J^y_{i,j-1}$.
When these four expressions are added, we arrive at
\begin{equation}
  J^{\mathrm{out}}_{ij} = \frac{4 \rho \phi_{ij}}{h (\rho^2 + R^2 B^2)}
  + \text{terms independent of $\phi_{ij}$}
  \label{eq:Jout}
\end{equation}
Now, if we replace the value of $\phi_{ij}$ by
\begin{equation}
  \phi^{\mathrm{new}}_{ij} =
  \phi_{ij} - J^{\mathrm{out}}_{ij} h (\rho^2 + R^2 B^2) / (4 \rho)
  \label{eq:fixphi}
\end{equation}
the right-hand side of Eq.~(\ref{eq:Jout}) will vanish.
We use Eq.~(\ref{eq:fixphi}) and similar ones with suitable modification
for grid points on the boundary, to update $\phi_{ij}$.

\section{Results}
\label{sec:results}

We conducted numerical computations on intrinsic indium antimonide
(InSb) semiconductor samples with the following properties near 300\,K.
\begin{equation}
  \begin{array}{r@{\;=\;}l@{\;}l}
    \rho & 8.0 T^{-5.333} \times 10^{8}
      & \mathrm{\Omega \, m} \\
    R & (-5.6 e^{-0.034 (T - 273)} - 0.9) \times 10^{-4}
      & \mathrm{m^3 \, A^{-1} \, s^{-1}} \\
    \alpha & (-3.2 + 0.01 (T - 273)) \times 10^{-4}
      & \mathrm{V \, K^{-1}} \\
    N & (-5.7 e^{- (T - 273) / 65} - 3.2) \times 10^{-5}
      & \mathrm{m^2 \, K^{-1} \, s^{-1}} \\
    M & 5 \times 10^{-2} & \mathrm{m^2 \, V^{-1} \, s^{-1}} \\
    \kappa & 1.4 \times 10^5 T^{-1.65} & \mathrm{W \, K^{-1} \, m^{-1}}
  \end{array}
\end{equation}
These values are not meant to be good fits to measurements.  They are
only shown here as example inputs to our code.  (In fact, $\rho$,
$\alpha$, and $N$ are rough fits to measurements near 300\,K by Ikeda
and others~\cite{Ikeda1,Ikeda2,Nakamura}, but the conditions are not
uniform: $N$ is measured under 4 Tesla, whereas $\alpha$ under no
magnetic field.)

On the basis of these values, we computed magnetoresistance and
Seebeck and Nernst effects with various sample geometry.

\begin{figure}
  \begin{center}
    \includegraphics[width=8cm]{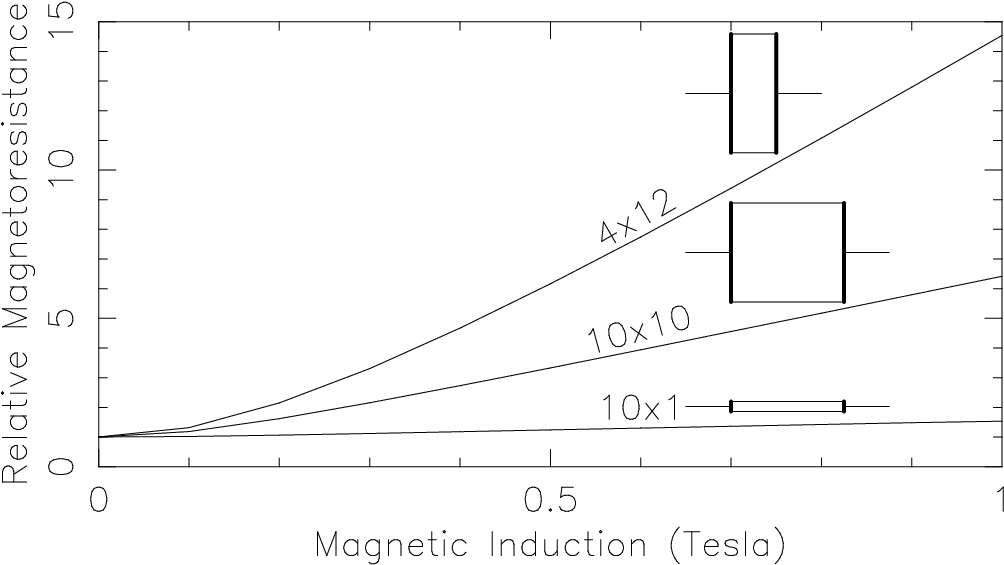}
    \caption{Simulation results of magnetoresistance for
      intrinsic InSb semiconductor with different geometry:
      length ($x$-direction) $\times$
      width ($y$-direction) $=$
      10\,mm $\times$ 1\,mm,
      10\,mm $\times$ 10\,mm,
      and 4\,mm $\times$ 12\,mm,
      with negligible thickness ($z$-direction).
      External magnetic field is along the $z$-direction.}
    \label{fig:hall}
  \end{center}
\end{figure}

The magnetoresistance results (Fig.~\ref{fig:hall}) are in good
agreement with experiment (\cite{WelkerWeiss}, Fig.~4.8 of
Seeger~\cite{Seeger}).

Seebeck and Nernst coefficients should vary much with the magnetic
field, so the results for these coefficients are to be compared with
those of different geometry under the same magnetic field.

\begin{figure}
  \begin{center}
    \includegraphics[width=8cm]{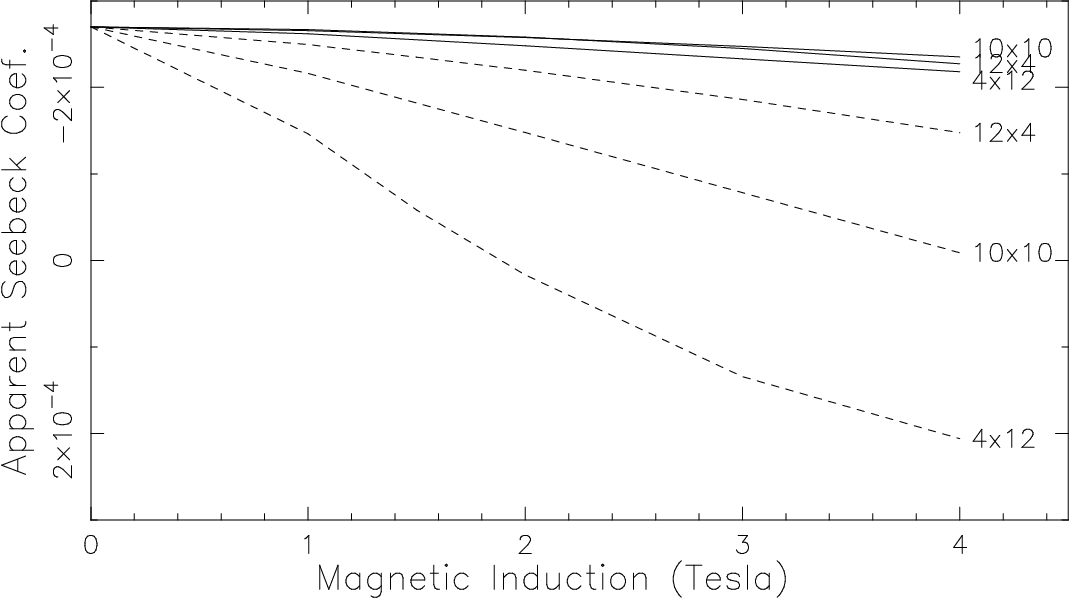}
    \caption{Simulation results of apparent Seebeck coefficient
      for intrinsic InSb samples with different geometry.  Apparent
      Seebeck coefficient is defined to be $\alpha_\mathrm{eff} =
      \Delta_x \phi / \Delta_x T$, where $\Delta_x \phi$ is
      longitudinal potential difference, and $\Delta_x T$ is
      longitudinal temperature difference.  Solid lines: Current leads
      are of negligible widths, Dashed lines: Current leads are as
      wide as the sample widths.}
    \label{fig:appsee}
  \end{center}
\end{figure}

As Fig.~\ref{fig:appsee} shows, the Seebeck effect is not sensitive to
geometry if current leads that measure longitudinal voltage difference
are narrow enough.  On the other hand, if current leads are as wide as
the sample width, Seebeck effect degrades and can even change sign for
short samples.  This tendency explains some experimental evidence for
the size dependence of magneto-Seebeck effect~\cite{Ertl}.

\begin{figure}
  \begin{center}
    \includegraphics[width=8cm]{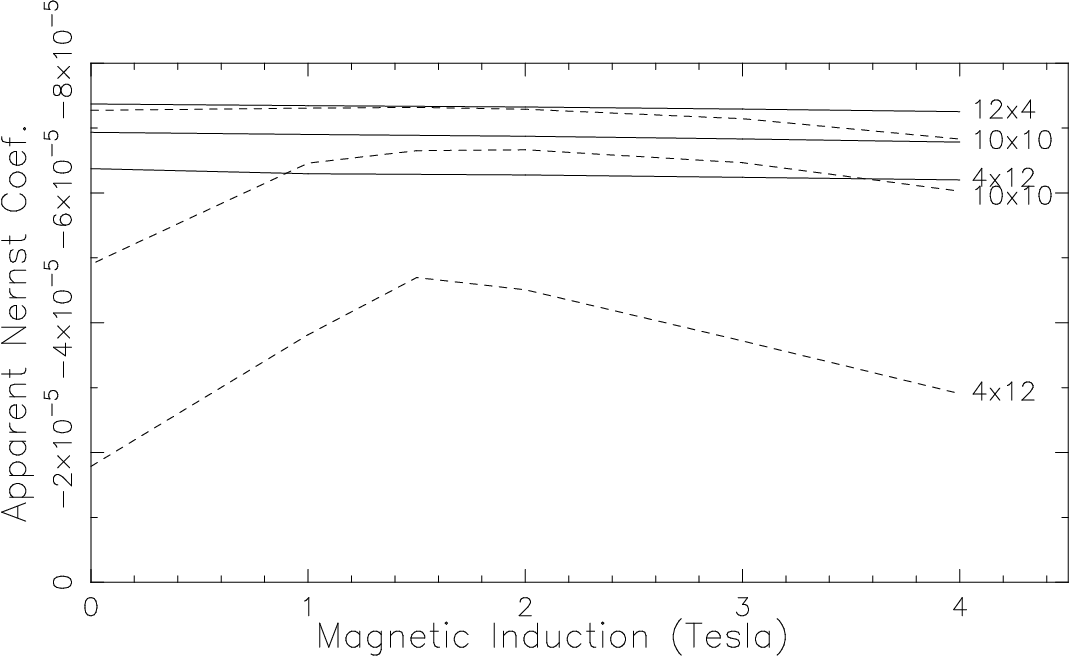}
    \caption{Simulation results of apparent Nernst coefficient
      for intrinsic InSb samples with different geometry.  Apparent
      Nernst coefficient is defined to be $N_{\mathrm{eff}} = L
      \Delta_y \phi / B W \Delta_x T$, where $\Delta_y \phi$ is
      transverse potential difference, $\Delta_x T$ longitudinal
      temperature difference, $L$ length (in $x$-direction), $W$ width
      (in $y$-direction), and $B$ magnetic induction.  Solid lines:
      Current leads are of negligible widths, Dashed lines: Current
      leads are as wide as the sample widths.}
    \label{fig:appner}
  \end{center}
\end{figure}

Similar tendency can be seen in Fig.~\ref{fig:appner} for the Nernst
effect.  In this case, however, geometry effect exists even with
narrow current leads.  This is because the Righi-Leduc effect causes
transverse ($y$-direction) temperature gradient that is proportional
to the magnetic field $B$.  This transverse temperature gradient in
turn causes transverse voltage gradient by the Seebeck effect.  When
divided by $B$, this transverse voltage gives a nearly constant bias
to the apparent Nernst coefficient.

This effect partly explains what Ikeda \textit{et al}~\cite{Ikeda1}
found out: Under 4 Tesla the apparent Nernst coefficient for their
``fat bridge'' (wide with legs) sample (Fig.~\ref{fig:fat}) is about
12 percent smaller than those of narrower one (Fig.~\ref{fig:thin}),
whereas our simulation gives only 7 to 8 percent smaller coefficient.
Though they are careful to make their current leads narrow, inevitable
finite widths of the leads might explain further effect.

\begin{figure}
  \begin{center}
    \includegraphics[width=8cm]{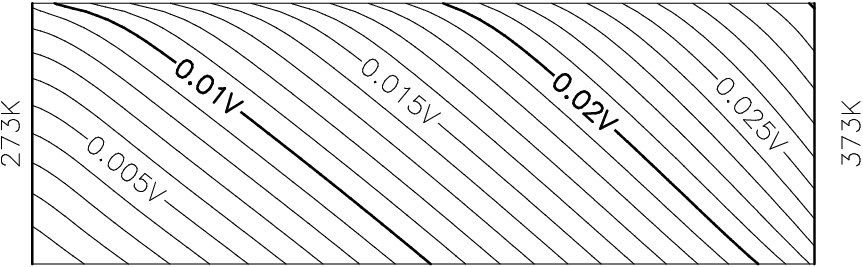}\\
    \includegraphics[width=8cm]{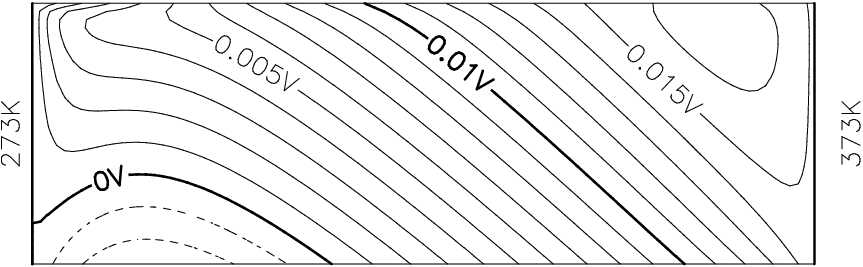}
    \caption{Contour maps of calculated potential distribution for a
      12\,mm $\times$ 4\,mm sample in 4-Tesla magnetic field.  Upper:
      Current leads are of negligible width.  Lower: Current leads are
      as wide as the sample width.  The left and the right edges are
      kept at 273\,K and 373\,K, respectively.}
    \label{fig:12x4}
  \end{center}
\end{figure}

\begin{figure}
  \begin{center}
    \includegraphics[width=80mm]{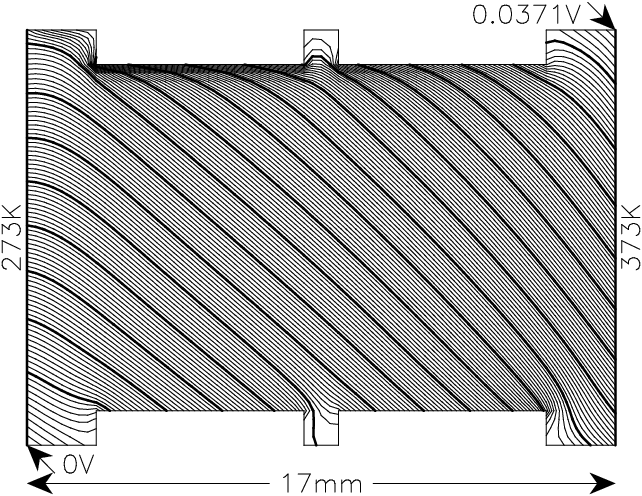}
  \end{center}
  \caption{Contour map of calculated potential distribution
    for ``fat bridge'' sample of Ikeda \textit{et al}~\cite{Ikeda1}.}
  \label{fig:fat}
\end{figure}

\begin{figure}
  \begin{center}
    \includegraphics[width=80mm]{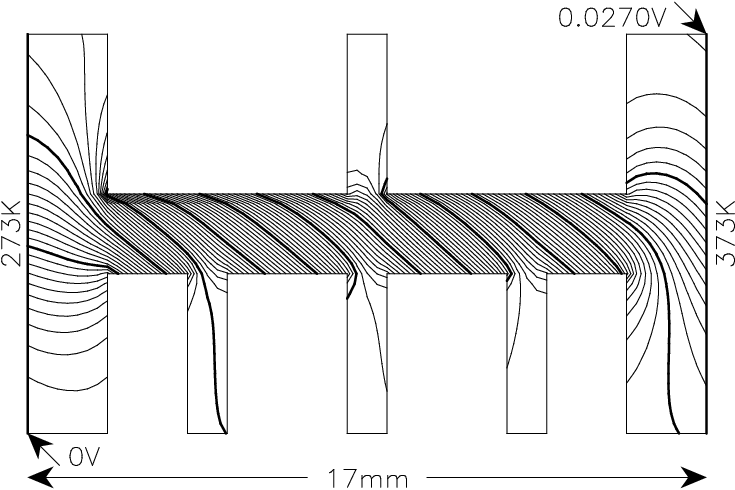}
  \end{center}
  \caption{Contour map of calculated potential distribution
    for ``bridge'' sample of Ikeda \textit{et al}.}
  \label{fig:thin}
\end{figure}

\section{Discussion}
\label{sec:discussion}

As can be seen from the contour maps, the potential distributions of
thermomagnetic samples are rather complicated.  They change
dramatically with different sample geometry.  Moreover, if we attach
current leads of finite widths to the cold and hot edges, much of the
transverse voltage gradient is shorted out, resulting in quite a
different potential distribution.  To correct for such a bias, careful
numerical calculation is necessary.

Inspection of Fig.~\ref{fig:12x4} (upper) and Fig.~\ref{fig:fat} shows
that the best way to generate electricity from wide samples under
magnetic field is to attach current leads to the bottom left and the
top right corners.  In this way, we can utilize both the Seebeck and
the Nernst effects.

Figs.~\ref{fig:fat} and \ref{fig:thin} also show that at the far end
of a small ``leg,'' potential gradient along the boundary decreases by
an order of magnitude or more, thus making measurement of Nernst
coefficients less prone to the widths of current leads.

\end{document}